\documentstyle[11pt,paspconf]{article}
\input psfig.tex

\begin{document}

\title{Numerical Simulations Of Advective Flows Around Black Holes}

\author{Sandip K. Chakrabarti\altaffilmark{1}, D. Ryu\altaffilmark{2}, D. Molteni\altaffilmark{3}, H. Sponholz\altaffilmark{4}, G. Lanzafame\altaffilmark{5}, G. Eggum\altaffilmark{6}}

\affil{1. Tata Institute Of Fundamental Research, Mumbai, 400005, INDIA}
\vspace{-0.2cm}
\affil{2. Chungnam National University, Daejeon, SOUTH KOREA}
\vspace{-0.2cm}
\affil{3. Istitut di Fisica, Via Archirafi 36, 90123 Palermo, ITALY}
\vspace{-0.2cm}
\affil{4. University of Kentucky, Lexington, USA}
\vspace{-0.2cm}
\affil{5. Osservatorio di Catania, Catania, Sicily, ITALY}
\vspace{-0.2cm}
\affil{6. Los Alamos National Laboratory, Los Alamos, NM, USA}

\vspace{0.5cm}

Observational results of compact objects are best understood using
advective accretion flows (Chakrabarti, 1996, 1997). We present here 
the results of numerical simulations of all possible types of such flows.

{\parfillskip=0pt Two parameter (specific energy ${\cal E}$ and specific angular momentum $\lambda$)
space of solutions of inviscid advective flow is classified into
`SA' (shocks in accretion), `NSA' (no shock in accretion),
`I' (inner sonic point only), `O' (outer sonic point only) etc. 
(Fig. 1 of Chakrabarti, 1997 and references therein). 
Fig. 1a shows examples of solutions (Molteni, Ryu \& Chakrabarti,
1996; Eggum, in preparation) from `SA', `I' and `O' regions where we superpose
analytical (solid) and numerical simulations (short
dashed curve is with SPH code and medium dashed curve is
with TVD code; very long dashed curve is with explicit/implicit code).
The agreement is excellent. In presence of cooling effects, 
shocks from `SA' oscillate (Fig. 1b) when the cooling timescale
roughly agrees with postshock infall time scale (Molteni, Sponholz \& Chakrabarti, 1996). 
The solid, long dashed and short dashed curves are 
drawn for $T^{1/2}$ (bremsstrahlung), $T^{0.4}$
and $T^{0.75}$ \par}
\vspace {6.6cm}
\noindent{\small Fig. 1: (a) Numerical simulations and theoretical
calculations of shock-free (O, I) and shocked (SA) flows. (b) Oscillation
of shocks when cooling effects are included.}
\newpage
\ \ \ \ \ 

\noindent cooling laws respectively.  
In the absence of steady shock solutions, shocks for parameters 
from `NSA' oscillate (Fig. 2) 
even in the absence of viscosity (Ryu et al. 1997). 
The oscillation frequency and amplitude roughly 
agree with those of quasi-periodic oscillation of black hole candidates. 
When the flow starts from a cool Keplerian disk,
it simply becomes sub-Keplerian before it enters through the horizon. 
Fig. 3a shows this behaviour where the ratio of $\lambda/\lambda_{Keplerian}$ is
plotted. When the flow deviates from a hot Keplerian
disk, it may develop a standing shock as well (Fig. 3b) (Molteni et al. 1996).\hfill\break

\vspace{1.7cm}
\noindent{\small Fig. 2: Oscillation of shock location \\
when the flow has two sonic points \\
but a steady shock condition is not satisfied.}
\vspace{1.7cm}

\vspace{6.2cm}

\noindent{\small Fig. 3: (a) Ratio $\lambda_{disk}/\lambda_{Kep}$ as obtained
from a 1D numerical simulation. (b) Another 1D simulation which shows formation
of sub-Keplerian flows and shocks from a hot Keplerian disk.}

\noindent To appear in: `Accretion Phenomena and Related Outflows', proceedings
of 163rd IAU Symposium, July-1996, Eds. D. Wickramsinghe, L. Ferrario
and G. Bicknell.

\noindent SKC's address AFTER November 26th, 1996:\\

\noindent Prof. S.K. Chakrabarti\\
\noindent S.N. Bose National Center for Basic Sciences\\
\noindent JD Block, Sector -III, Salt Lake\\
\noindent Calcutta 700091, INDIA\\

\noindent e-mail: chakraba@bose.ernet.in  OR chakraba@tifrc2.tifr.res.in \\ 

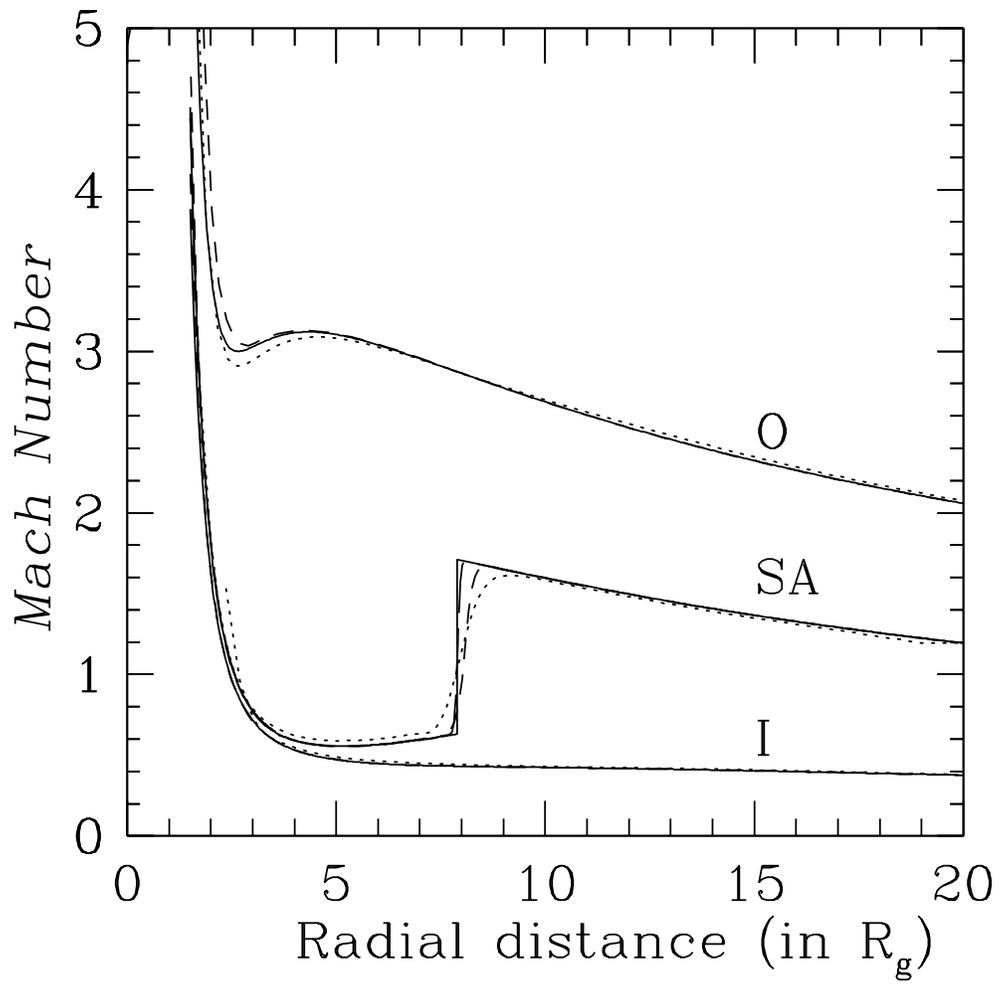
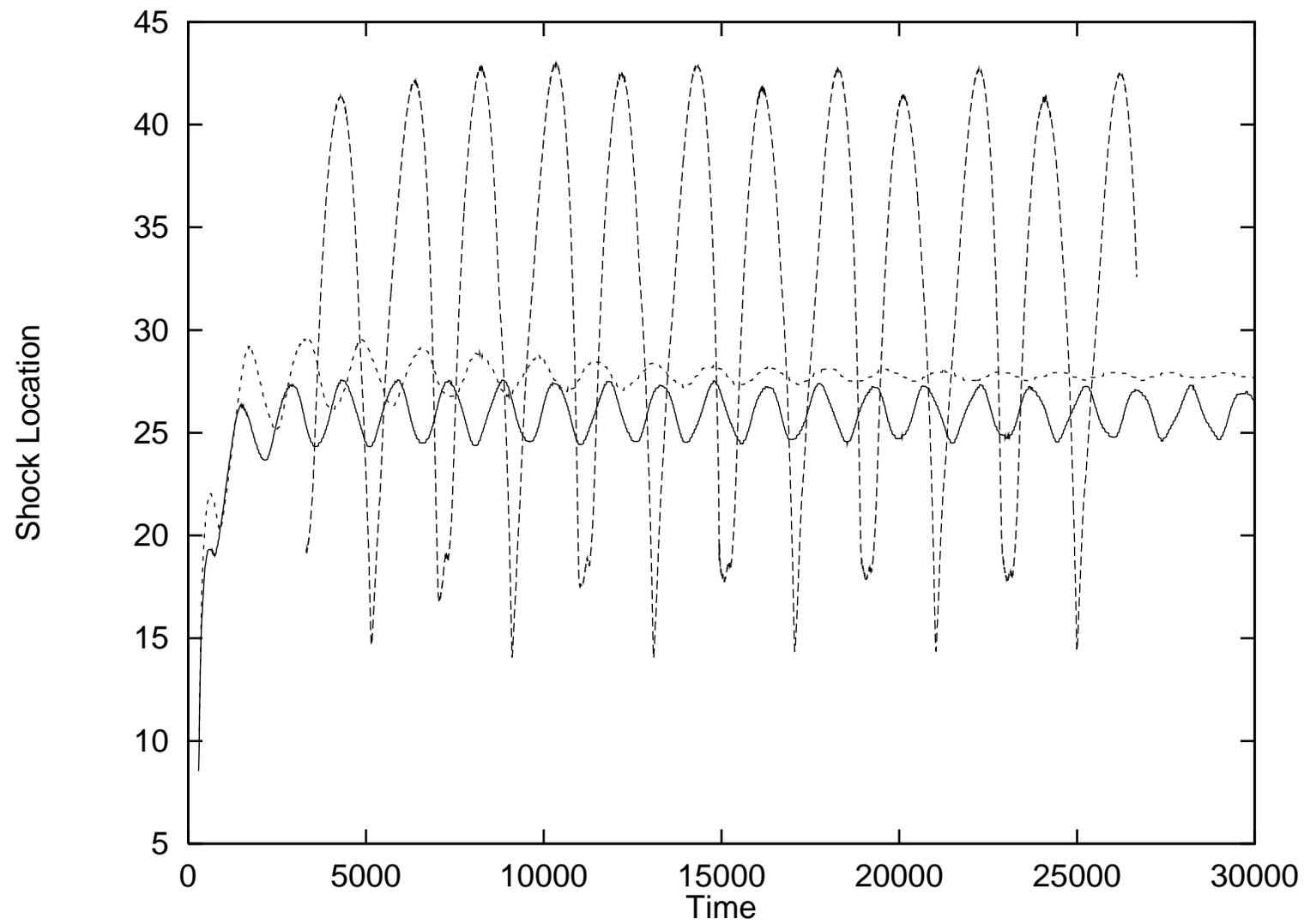
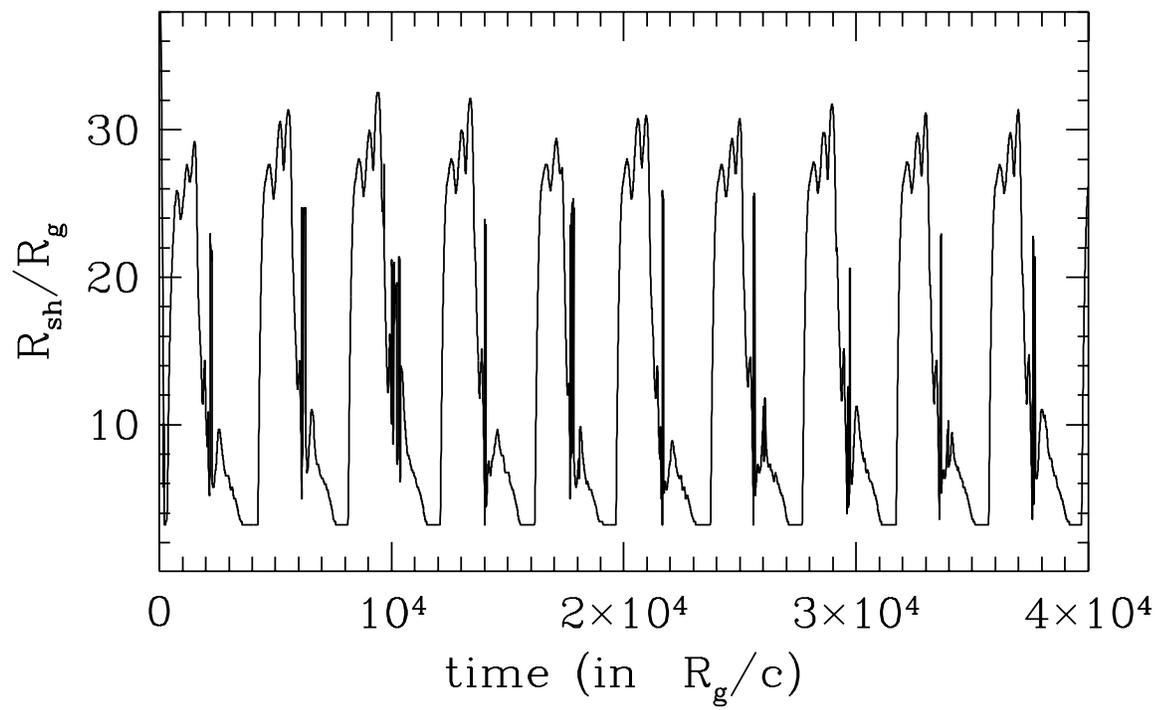
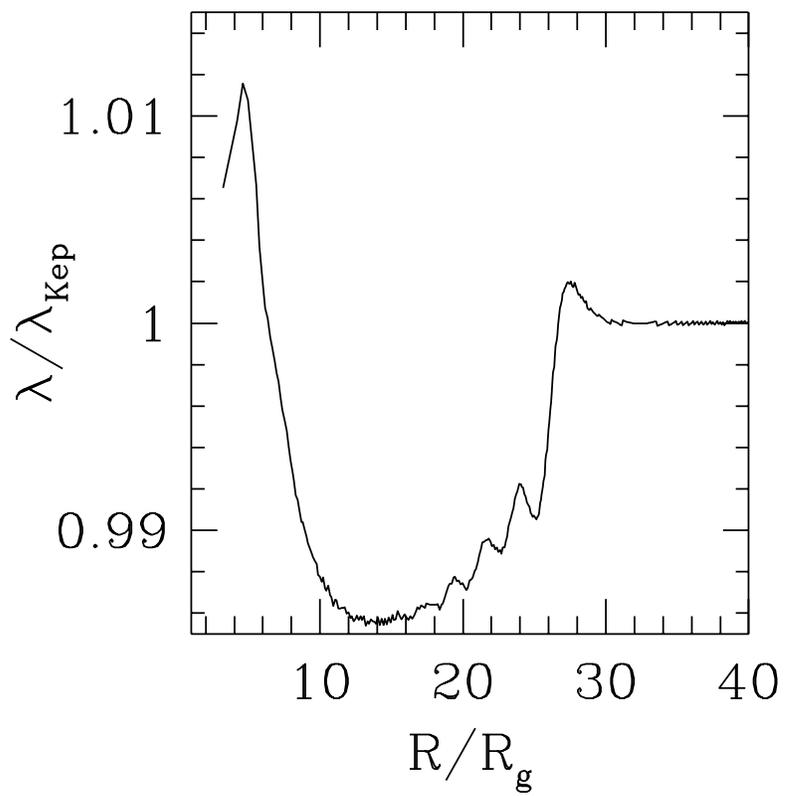
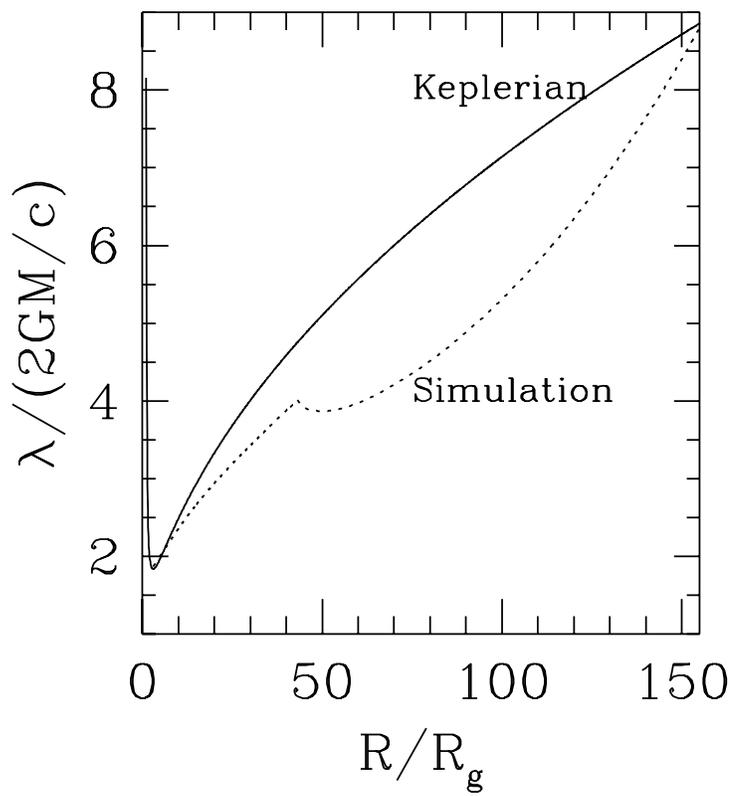
\end{document}